# A Method for Deriving Technical Requirements of Digital Twins as Industrial Product-Service System Enablers

Jürgen Dobaj[1] [0000-0001-6460-8080], Andreas Riel[2] [0000-0001-9859-019X], Georg Macher[1] [0000-0001-9215-3300] and Markus Egretzberger[3]

[1]Graz University of Technology, Graz, Austria, `juergen.dobaj@tugraz.at`
[2]Grenoble Alps University, Grenoble INP, G-SCOP, CNRS, Grenoble, France
[3]Andritz Hydro GmbH, R&D Automation, Vienna, Austria

**Abstract.** Industrial Product-Service Systems (IPSS) are increasingly dominant in several sectors. Predominant value-adding services provided for industrial assets such as production systems, electric power plants, and car fleets are remote asset maintenance, monitoring, control, and reconfiguration. IPSS designers lack methods and tools supporting them in systematically deriving technical design requirements for the underlying Cyber-Physical System (CPS) IPSS services. At the same time, the use of Digital Twins (DTs) as digital representations of CPS assets is becoming increasingly feasible thanks to powerful, networked information technology (IT) and operation technology (OT) infrastructures and the ubiquity of sensors and data. This paper proposes a method for guiding IPSS designers in the specification and implementation of DT instances to serve as the key enablers of IPSS services. The systematic mapping of the continuous IT design-build-deployment cycle concept to the OT domain of CPS is at the heart of the applied methodology, which is complemented by a stakeholder-driven requirements elicitation. The key contribution is a structured method for deriving technical design requirements for DT instances as IPSS. This method is validated on real-world use cases in an evaluation environment for distributed CPS IPSS.

## 1 Introduction

According to a very recent analysis by Brissaud et al. [1], the Industrial Product-Service-System (IPSS) use cases considered most relevant at present and in the near future are remote asset maintenance and remote asset monitoring, as well as remote asset control and reconfiguration. Cyber-physical systems (CPS) that are characterized by a high level of connectivity, configurability, adaptability, and flexibility during operations are essential elements of Industry 4.0. They bring along new challenges for assuring the availability and integrity of industrial assets at any time. As virtual counterparts of such assets, Digital Twins (DTs) present an opportunity for shifting essential parts of the effort of predicting, evaluating, and preparing asset configurations during design time rather than during expensive and availability-critical operation time. From a design perspective, however, there is no structured methodology supporting IPSS designers in deriving service design requirements and mapping them to DT design and deployment requirements. As for the former challenge, Pezzotta et al. [2] propose a Service Engineering Methodology (SEEM) to guide corporate decision-makers in defining the (I)PSS offering that serves their strategic needs and expectations. They suggest a structured decision-making process to (i) define the PSS offering most aligned with company products and customer needs, (ii) (re)engineer the (existing) service delivery processes, and (iii) balance the external performance (e.g., customer satisfaction, delivery time, service cycle time) with the internal performance (i.e., efficiency) of the service delivery process. The work presented here goes far beyond SEEM in that it proposes a systematic, structured approach that facilitates the derivation of technical design, deployment, and operation requirements for DTs as IPSS services enablers. It achieves this through integrating the IPSS business and IPSS provider needs into the CPS IPSS design, development, and operation processes.

The remainder of this work is structured as follows. Section 2 explains related work and the research methodology. Sections 3 to 5 provide the core contribution of this work. As a first step, Section 3 describes our SEEM extension for the stakeholder-driven derivation of technical CPS and DT design requirements. Section 4 establishes a systematic mapping of the continuous IT design-build-deployment cycle concept to the operational domain of CPS. Based on that, Section 5 introduces our structured method for the technical design of CPS and DT elements. Section 6 elaborates on the validation of the contributions. Finally, Section 6 concludes this work.

## 2      Related Work and Methodology

Meierhofer et al. are among the rare to propose a holistic and actionable concept for modeling industrial service ecosystems based on the DT [3]. Their key proposal is to switch from the classical Goods Dominant Logic (GDL) driven design process to a Service-Dominant Logic (SDL) driven design process. Schuh et al. [4] focus on maintenance, repair, and overhaul (MRO) services for machine manufacturers. The paper maps the MRO services and their single elements through a case study, which sets the basis for designing the digital shadow. Stark et al. [5] investigate methodological, technological, operative, and business aspects of developing and operating Digital Twins. Very recently only, Leng et al. [6] presented a DT-driven approach to the design-time determination of expected lower-level CPS reconfiguration costs during planned downtime. Abromocici et al. [7] propose a purely conceptual model for product reconfiguration in its use phase. In their paper, they envision virtual product twins for managing smart product reconfiguration processes. In a different application domain, Zhou et al. [8] propose a DT framework for power grid online analysis. Their focus is on high-performance computing of the DT of the national grid energy management system for real-time tracking. However, none of these and similar works we have found deals with the research question of *how to systematically derive requirements for establishing a seamless link between the design of the service-enabling DTs and those for their (continuous) deployment, maintenance, operation, and optimization in the real CPS IPSS environment.*

The methodology to address this question is the following: first, SEEM is extended to also include the business- and provider-focused viewpoints on IPSS design rather than taking the classically pure customer-oriented one. Applying this SEEM extension to the customer, business, and provider needs derived from literature [1] and during our action research activities with our industry partner leads to functional and non-functional requirements for the IPSS services and their enabling DTs.

Our hypothesis is that these requirements can be satisfied through a mapping of the DevOps lifecycle from the IT domain [9] to CPS-based IPSS. This mapping allows establishing a structured and layered concept that guides designers in the systematic derivation of functional and technical DT design, deployment, operation, monitoring, and reconfiguration requirements based on the IPSS (service) needs within and across their individual lifecycle phases. For validation, we applied our method for the requirements derivation of a representative real-world use case that we investigated in our action research activity with Andritz Hydro GmbH. The technical requirements and implementation aspects including the use case evaluation are presented in [10]. In the work presented here, we first introduce our proposed design method and second demonstrate its applicability to derive DT (and CPS) requirements for this IPSS use case.

## 3  Extending SEEM for Technical Requirements Derivation

Figure 1 shows the SEEM extension we propose, applied to our IPSS case study. Our approach can be structured into three major phases: use case design, IPSS design, and CPS design. First, use case design focuses on the identification of the stakeholders' needs and wishes (see Figure 1 steps 1 to 3). These stakeholder demands are integrated into the design of use cases that describe the stakeholders' interaction with the IPSS.

Second, IPSS design focuses on the service offering to these stakeholders (see steps 4 and 5). In particular, it focuses on the design of interfaces and activities that shall enable stakeholders the service-oriented access to functions and capabilities of the product that is underlying the IPSS. The outcome shall be a series of (sequential and parallel) activities describing how specific stakeholder IPSS requirements are realized.

The third and last phase (i.e., the CPS design in the steps 6 to 7) focuses on the technical realization of individual activities. In this phase, the identified activities serve as an input for the design of the functional CPS IPSS concept in step 6, subsequently also denoted as functional IPSS prototype. Step 7 transforms this functional concept into technical requirements that determine the underlying CPS functional, behavioral, and structural design requirements.

Depending on the industrial sector and the specific use cases, it is necessary to consider domain-specific standards and engineering processes in the CPS design phase to ensure, e.g., system safety and security [11,12]. In addition, fundamental factors like cost, performance, and dependability, as well as other system properties such as usability, manageability, and adaptability, influence technical system design. A detailed consideration of these issues is beyond the scope of this work. The remainder of this work focuses on the technical design of DT elements that operate in the context of CPS IPSS.

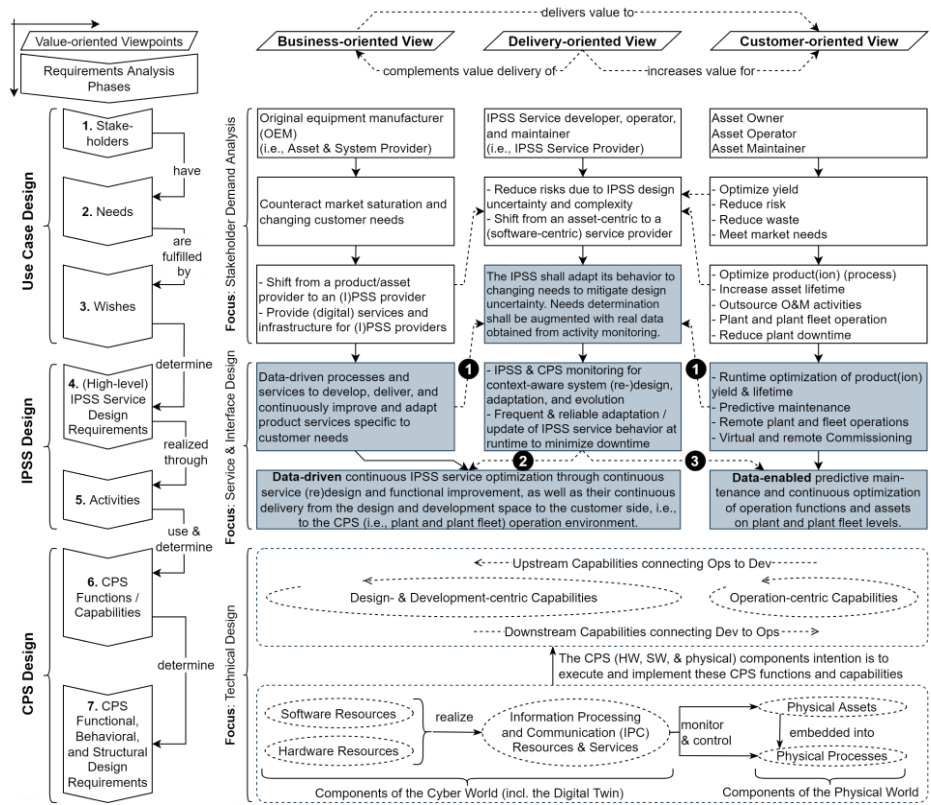

**Fig. 1.** Our SEEM extension for integrated stakeholder-driven IPSS requirements analysis.

Although the focus is on DT element design, all presented methods are generically applicable and can be tailored for the design of any digital (software-driven) CPS element.

### 3.1 The Potential of an Integrated View on Multiple Stakeholders' Needs

While SEEM has a strong focus on the customer-oriented requirements path (right column in Figure 1) and its subsequent mapping to company resources, our extension to SEEM is twofold. First, we complement the customer requirements with the OEM's top-level business objectives (left column), and second, we integrate IPSS provider requirements for the development, operation, and maintenance of (digital) IPSS services and their underlying CPS elements, including the said DT elements (middle column).

In particular, the three main phases of our approach are structured into seven substeps represented by the rows in Figure 1. Each cell in Figure 1 summarizes the needs, wishes, and requirements we identified for the different stakeholders in our case study. When creating such a requirements tree structure, the different requirements paths (i.e., stakeholder viewpoints) do not only lead to disjoint requirements. Instead, the paths may also lead to similar and contradicting requirements. Our experience showed that (graphical) relationships established between needs, wishes, and requirements across stakeholder viewpoints have the potential (a) to support the subsequent prioritization of functional CPS requirements and thereby balancing contradictory stakeholder demands, (b) to reduce stakeholder risks through risk sharing, and (c) to facilitate the identification of innovative IPSS use cases and service offerings. Next, we discuss some relationships that are indicated in Figure 1 by the dashed lines.

### 3.2 Use Case Design meets IPSS Design: An Integrated View on Stakeholder Demands and IPSS Requirements

**Business View.** In our case study, the use case analysis revealed that the OEM needs to adapt its business strategy to market trends like market saturation in developed countries and the increasing end-customer expectations for, e.g., flexibility, customization, and pay-per-use. These market trends are all drivers of product servitization [1,13,14]. Consequently, the OEM's aim is to provide IPSS services for the (remote) operation, maintenance, and optimization of assets and processes that are embedded into the CPS that underlies the IPSS. In our case study, this means that these services shall be integrated on hydropower plant (HPP) and HPP fleet levels and provided to customers (i.e., asset owners, operators, and maintainers) for brownfield and greenfield deployments. Besides servitization, our analyses showed that said customers frequently wish to outsource their operations and maintenance (O&M) activities. This is another business opportunity for the OEM that is also wishing to transition from a pure asset and system provider to an IPSS provider.

**Customer View.** As for the customer requirements, we focused on those Brissaud et al. [1] identified to be representative of the most frequent IPSS today and in the near future, including remote asset maintenance and monitoring services, as well as remote asset reconfiguration and operation services. These services rely on Model Predictive Control (MPC), frequently facilitated by data-intensive machine-learning algorithms like Deep Neural Networks (DNN). DTs as virtual counterparts to these assets are accepted as an enabler for IPSS in general and for the above use cases in particular [3–7].

**Delivery View.** To deliver the said DT-enabled IPSS offering, mainly software-driven information processing, and communication (IPC) functions and capabilities need to be integrated into CPS design. These IPC functions pursue the goal of enabling stakeholders service-oriented access to CPS asset and process capabilities. The middle column in Figure 1 integrates the delivery-oriented aspects related to these IPC functions into our stakeholder analysis. In particular, an IPSS provider is responsible for the development, delivery (i.e., deployment and commissioning), operation, reconfiguration, and maintenance of the IPSS enabling technologies and functions, i.e., the DT and IPC elements that are embedded into the CPS.

Our analysis of the said delivery-oriented aspects revealed that IPSS providers face high planning (and hence implementation and operation) risks due to design uncertainty [15]. These risks are related to unclear requirements of IPSS service behavior, operation environment, and emergent CPS behavior that includes taking pace with changing customer needs, emerging cyber security attacks, the increasing risk of fatal (software) bugs in complex and interconnected CPS, as well as the evolution of standards and technology throughout the entire CPS lifetime [11,16,17]. These findings are also supported by industry studies [1,18,19], concluding that the risk imposed by technical, user behavioral, and service provisioning uncertainties is evident and discourages the industry from adopting the IPSS concept. To address said IPSS provider risks, we defined the wish that the IPSS shall be able to adapt its behavior (i.e., its activities and thereby its service offering) to changing needs and that the need identification shall be augmented with real IPSS and CPS operation data.

**Integrated View.** The grey boxes in Figure 1 highlight the essential stakeholder requirements of our IPSS case study, which can be summarized by the need to create an adaptable and evolvable IPSS. The realization of such an adaptable IPSS increases system complexity and hence system development cost, which, however, shall be kept minimal. Nevertheless, the integrated view on stakeholder demands, see Figure 1 label ❶, reveals that IPSS adaptability is essential to realize business and customer IPSS requirements alike: (a) To meet business requirements, the IPSS shall monitor user (i.e., customer) behavior for the data-driven design and development of products and product services. (b) To meet customer requirements, the IPSS shall monitor CPS asset and process operations for data-enabled predictive maintenance and the continuous (runtime) optimization of CPS O&M functions. As indicated by Figure 1 label ❶, the said monitoring capabilities are also required by the IPSS provider for context-aware system (re-)design, adaptation, and evolution. In addition to that, an IPSS provider requires services for the frequent and reliable adaptation and update of IPSS service behavior. Such adaptations shall be possible at runtime (i.e., during full CPS IPSS operation) to minimize CPS downtime.

**Summary.** Summing up, the delivery-oriented view complements business value-delivery (see Figure 1 ❷) by supporting (a) data-driven activities that monitor and analyze system and stakeholder behavior to provide the analytical capabilities for evidence-based optimization and adaptation of the IPSS offering to unanticipated, changing customer, provider, and business needs. Thereby the delivery-oriented activities increase value for the customer (see Figure 1 ❸) through supporting (b) data-enabled activities that facilitate the context-aware continuous (runtime) adaptation and optimization of IPSS services and CPS functions.

**Conclusion.** The identified IPSS requirements are key characteristics of (Smart) IPSS [1] and DevOps [20] alike. Both concepts heavily rely on data (i.e., information about their essential processes, activities, assets, and services) and require closed-loop design and context-awareness as essential design characteristics throughout all their lifecycle phases [10,21]. We propose that the integration of DevOps principles into the IPSS lifecycle is essential to cope with system complexity and emergent, unforeseen system behavior and requirements (i.e., design uncertainty). Furthermore, we propose that the DT concept is suitable to enable IPSS and DevOps activities alike. Hence, integrated CPS and DT element design has the potential to build (digitally driven) solutions that co-create value for all stakeholders.

### 3.3 CPS Design: From IPSS to Technical CPS and DT Design Requirements

In the CPS design phase, we propose an additional SEEM extension to transfer the identified IPSS requirements to technical CPS and DT design requirements. Instead of instantly mapping IPSS activities to company resources, our approach introduces a function-oriented abstraction layer (i.e., Figure 1 step 6). In this step, we use Sequence Diagrams (SDs) for the structured mapping of IPSS activities to CPS element functions and capabilities. The thereby obtained functional description of CPS elements and CPS element interactions represents the IPSS prototype. This prototype (a) serves as a reference model for early phase (simulation-based) IPSS use case validation and (b) serves as a functional system concept for the detailed technical CPS design in step 7.

In the course of Section 3, we proposed that DTs and DevOps are essential concepts that shall be integrated into IPSS design to meet stakeholder demands. Next, in Section 4, we describe these concepts to establish the relationships between DevOps, DTs, CPS, and IPSS. In Section 5, we integrate these concepts into our SD-based design approach.

## 4 Characterizing the DevOps Lifecycle for DT-enabled CPS IPSS

### 4.1 Characterizing DevOps

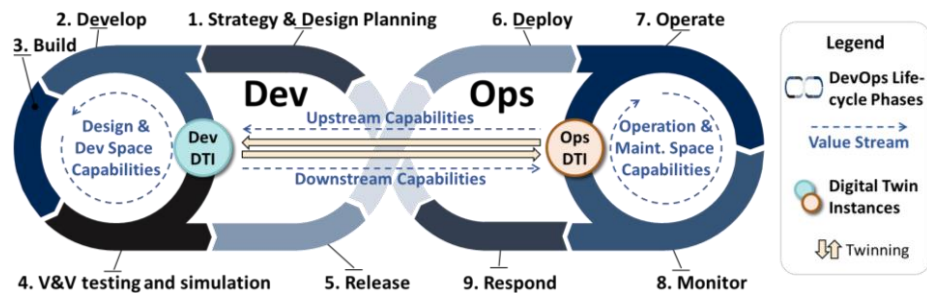

**Fig. 2.** DevOps-inspired lifecycle for DT-enabled IPSS services (modified from [9])

**DevOps.** DevOps, with its origin in the information technology (IT) sector, is a collection of principles from computer science aiming toward the tight integration and collaboration of IT development (Dev) and IT operations (Ops) activities and teams [9,20,22]. As shown in Figure 2, these DevOps activities can be structured into nine lifecycle phases that can be roughly described through four value streams [10]. These value streams group the activities and capabilities required to create value for stakeholders within and across specific DevOps lifecycle phases. Dobaj et al. [10] describe these value streams as follows: (a) The creation of value in the (Design &) Dev space by developing, building, and testing new and improved services. (b) The creation of value by the downstream delivery of these services from the Dev space to the operations environment for their operationalization. (c) The creation of value by using, maintaining, and monitoring the provided services in the Operation & Maintenance (O&M) space. (d) The creation of value by operations monitoring and the delivery of upstream feedback for a data-driven strategy and design planning in the Design (& Dev) space.

**Reducing Design Uncertainty with DevOps.** The tight integration and collaboration of Dev and Ops activities are essential for the closed-loop implementation of these four value streams to enables the desired continuous building, testing, and deployment of software services upon each software (design) modification. Consequently, the provided services can be continuously improved and adapted to changing needs, which finally reduces system design uncertainty. For example, particular key strengths of DevOps are the short, close-loop (upstream and downstream) feedback cycles between Dev and Ops [20], which are enabled through rigorous operations monitoring and the frequent and reliable deployment of software services changes during full system operation. While such a scenario is pertinent for IT environments, we must note that we are far from achieving the same for OT environments. Only very recently, we were the first to demonstrate the initial steps for integrating DevOps into CPS OT environments [10].

### 4.2 Characterizing DT-enabled DevOps for IPSS

**Digital Twin Elements.** Figure 3 (modified from [23]) introduces the DT concept. In particular, Figure 3 (a) shows the concept of twinning, where a real entity is twinned (or mirrored) from real space into virtual space and vice versa. Twinning is realized by the connection between real and virtual space, connecting a real entity to its virtual counterpart, i.e., the Real Twin (RT) with its Digital Twin (DT). Traditionally, the DT concept focused on the virtual representation of physical product properties throughout the product lifecycle phases [23]. Today, the DT concept is also used to mirror the lifecycle and properties of (virtual) software products [24,25] and (mixed) cyber-physical products [10,26]. To reflect these relationships, we modified Figure 3 (b) from [23], which shows DT elements (vertical axis) in the context of their associated RT (product) lifecycle phases as of [27].

**DT and RT Lifecycle.** The DT starts life as a Digital Twin Prototype (DTP) in the design phase. A DTP is for instance a digital model describing DT and/or RT properties such as structure and behavior. Digital Twin Instances (DTIs) are created for each RT during their realize phase (e.g., during manufacturing for physical products; and by deployment and execution for software products). A DTI conforms to its DTP and mirrors its associated RT properties in virtual space. The accumulation of DTIs forms a Digital Twin Aggregate (DTA) to provide more complex and averaging DT-enabled capabilities. Both the DT Instances and Aggregates exist within their Digital Twin Environment (DTE), which is the virtual representation of the environment within which the real product exists. The DT Instances, Aggregates, and Environments persist beyond the actual life of its real product that ends in the Retire/Dispose phase.

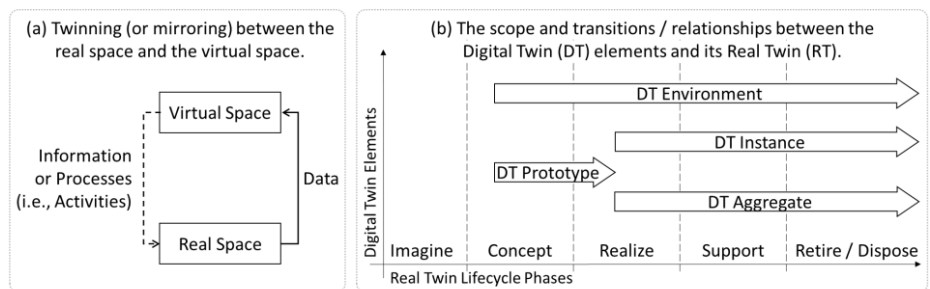

**Fig. 3.** (a) Twinning concept and (b) relations between DT and RT elements in their lifecycles.

From a business perspective, each product and each product-service are created for one dedicated purpose, which is the creation and delivery of value for stakeholders. Similar to these products and services (i.e., the RTs), each DTI is designed and created to optimize the value creation capabilities of its associated RT. As of Section 3, it is of utmost importance for all stakeholders to reduce design uncertainty, which as propose in Section 4.1, can be met by integrating DevOps principles into the IPSS lifecycle.

**DT-enabled IPSS DevOps Lifecycle.** To integrate DevOps principles into IPSS lifecycle activities, we propose that the four value streams must be integrated into the design of the underlying CPS lifecycle and its IT and OT functions. To that end, the CPS must support DevOps activities within and across (a) all CPS lifecycle phases and (b) all CPS layers. Section 4.3 characterizes these CPS layers for DT-enabled DevOps to explicitly define the DT Environments in the context of CPS IPSS.

From a lifecycle perspective, we propose that the DT concept can serve as enabler technology to realize the said value stream capabilities and functions, which we introduced by the DTIs in Figure 2 and make now more explicit in Figure 4. In this DevOps-inspired lifecycle for IPSS, the DTIs serve as a common knowledge base within and across the individual lifecycle phases to support the realization of overarching functions and activities within the four value streams. Since the individual value streams impose contradictory requirements on DTI usage and, therefore, also on their design and implementation, we recommend the separation of DTIs through a modular and service-oriented DTI design. Such a design shall enable the decoupled development, operation, and evolution of DTIs and their associated IT and OT functions. The aggregation of individual DTIs allows realizing overarching functionality.

For example, the Ops DTI is ideally deeply integrated into the control of the CPS (assets and processes) and shall therefore be designed for availability, integrity, safety, and real-time operation in resource constraint OT environments. To that purpose, the

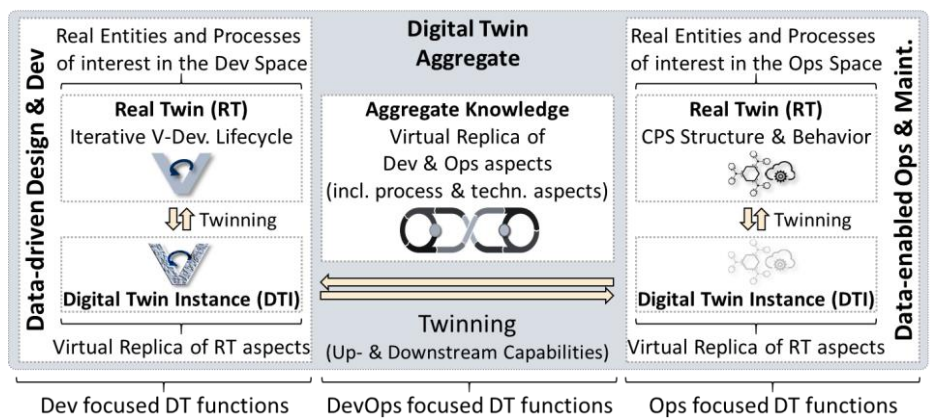

**Fig. 4.** Concept-level design: Digital Twin-enabled DevOps for CPS IPSS

Ops DTIs shall reflect aspects of the CPS structure and behavior to support the realization of data-enabled CPS and IPSS O&M activities (see Figure 4). The Dev DTI, instead, resides in the IT office environment for mainly data-driven system planning, testing, and simulation. Such IT services are less critical compared to OT services and can be designed for and executed in modern Cloud environments. It is the responsibility of the DTA to aggregate the development and operation knowledge to establish the context-awareness for realizing the up- and downstream capabilities to ensure data, context, function, and capability sharing across DTIs and DevOps phases. In Section 5, we extend on the conceptual system design shown in Figure 4, where we divide the Dev and Ops DTIs into sub-DTIs and define the communication between these sub-DTIs.

### 4.3 Characterizing the DT Environment within CPS IPSS

All views (a-d) that are shown in Figure 5 are vertically aligned to the Automation Pyramid (AP) levels [28]. The AP is an architectural concept that describes the typical structure and organization inherent to large-scale networked industrial automation and control systems. The properties and needs of each AP level significantly influence the design and implementation of CPS IPSS functions and services. The vertical alignment to these levels shall highlight that all views are strictly related to each other and that individual design aspects within one view may influence the design and function of other views. The conceptual DT elements at the top of Figure 5 are associated with these views to show the relationships between DT and CPS element design.

**CPS IPSS Structure and Organization.** In particular, view (a) shows the typical, physical, hierarchical layout of the CPS structure, including its IPC resource clusters that execute the distributed and hierarchically organized (software-driven) functions shown in view (b). The implementation of these functions determines the CPS behavior and supports the realization of business, development, and operation activities. View (d) maps these activities to the four DevOps value streams. Finally, view (c) shows our proposed realization of the Dev DTI and the Ops DTI as of Figure 2. As of Figure 4, the aggregated knowledge of these DTIs creates the context-awareness across AP levels to realize DevOps activites. The communication between the distributed DTIs is established using IT NWs on the upper AP levels and OT NWs on the lower AP levels.

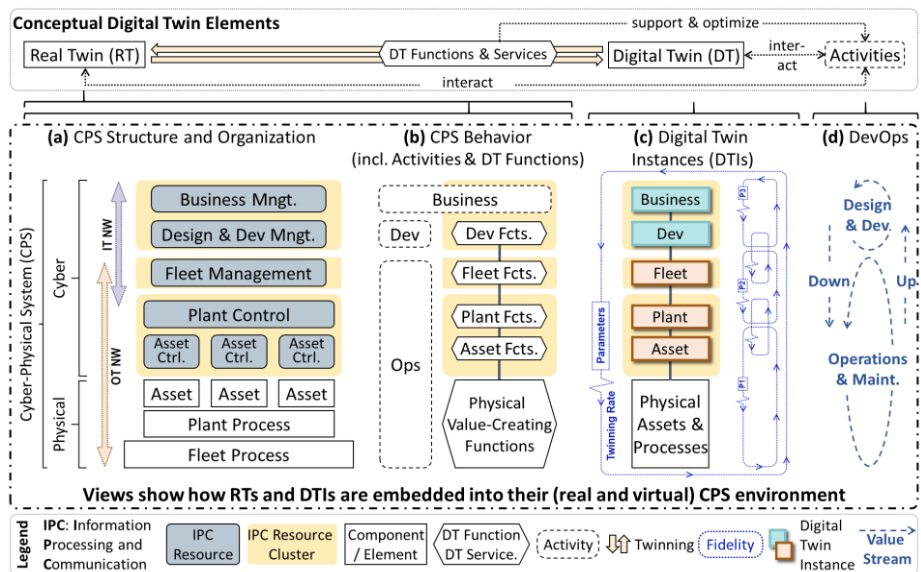

**Fig. 5.** Generic architecture models characterizing the relationships between the conceptual DT elements and their realization in their real and virtual (CPS IPSS) environments.

**IT NW.** The top layer is called the business or management level. At this level, the DTI provides key performance indicators (KPIs) of the overall CPS IPSS to drive business decisions. Below that, DTIs are used for strategic system operations and design planning, development, simulation, and verification & validation. Due to their functional aspects, the first two levels can be mapped to the Design & Dev value stream. Cloud IT solutions can be used for information processing and exchange at these levels.

**OT NW.** Fleet management and plant control are denoted as the supervisory level responsible for supervisory control and data acquisition (SCADA). The lowest Cyber-level is denoted as the control level, responsible for domain and asset control on the plant level. In particular, the programmable logic controllers (PLCs) are connected via the sensors, actuators, and Fieldbus networks to the assets and processes under control.

**CPS and DT Functions and Services.** As shown by view (b), CPS behavior is determined by the functions and services supported on the individual levels. These functions are executed on the distributed IPC resources to coordinate and control the physical assets and processes, to coordinate and implement DevOps and business activities, and to manage the IPC cluster resources and their usage.

**DT Environment.** The views in Figure 5 describe the DTE for the DT elements as of Figures 2 and 4. In particular, view (c) shows how DTIs are integrated into a typical CPS environment, where the Dev and Ops DTIs are composed of multiple DTIs that are distributed within and across the AP levels. The aggregated knowledge of these DTIs shall provide the necessary context-awareness for both the CPS and IPSS function and service composition, coordination, and management across layers and lifecycle phases. To that purpose, we propose a modular and service-oriented DTI design connected via said IT and OT NWs. The blue boxes surrounding the DTIs in view (c) indicate the twinning fidelity, which describes the twinning accuracy determined by the twinned parameters and the twinning rate of these parameters [23]. This twinning fidelity heavily depends on the use case and the individual (process, functional, and technological) needs of and across AP levels. To meet these needs, the twinning fidelity must be adapted between AP levels. Similar to the DevOps aspects, we integrate said DT Environment aspects into our proposed SD-based design approach in Section 5.

## 5      Deriving Technical Digital Twin Requirements: A Case Study

The aim of our case study is to implement DevOps principles within the IPSS lifecycle. For that purpose, DTIs shall manage and coordinate the DevOps activities within and across the IPSS lifecycle phases (see Figures 2 and 4). In Section 4, we characterize the relationships between CPS and DT elements (i.e., the DT Environment) for integrating DevOps activities into the CPS underlying IPSS. To support a structured mapping of these activities (see Figure 1 step 5) to the technical CPS design (see Figure 1 step 7) we added step 6 in Figure 1 to obtain a functional IPSS prototype. As introduced in Section 3.3, this prototype is a model that describes the CPS elements and their interactions for realizing the associated IPSS activities. The obtained functional aspects determine the subsequent technical design of the CPS functions, behavior, and structure in step 7. As also stated in Section 3.3, we propose to use Sequence Diagrams (SDs) to create that structured mapping in step 6.

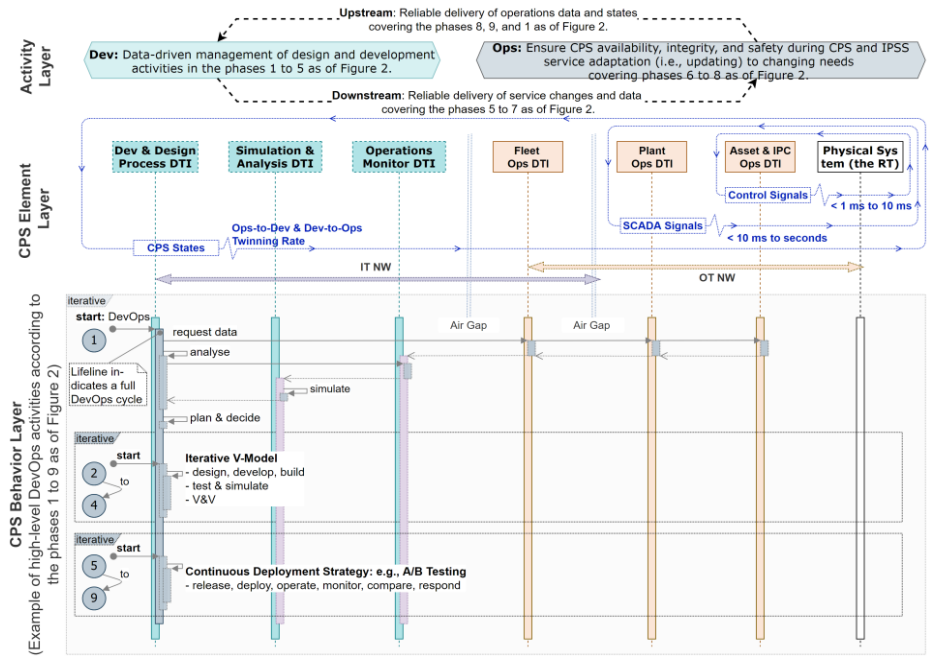

**Fig. 6.** Transforming IPSS activities into technical CPS (and DT) design requirements.

Figure 6 shows the proposed SD for our case study. The SD is structured into three layers: activity, CPS element, and CPS behavior layer. The activity layer on top shall structure the IPSS activities into the four DevOps value streams intended to make the relationships to the DevOps lifecycle explicit. The CPS element layer consists of CPS element instances that are required to support or implement an associated activity at the activity layer. To establish a proper mapping, the CPS element layer also follows the DevOps value stream structure. Once the mapping is established, CPS element requirements can be derived from DevOps activities and their associated Automation Pyramid (AP) level shown in Figure 5. The CPS behavior layer describes the interaction between CPS elements using the standard XML SD notation of messages.

In our case study, where we are specifically interested in the design of DT elements, these interactions explicitly describe the data exchange between different DT elements (i.e., CPS elements), which provides a structured way to determine the twinning fidelity requirements for the up- and downstream capabilities that span multiple AP and NW levels. In addition, the functions attached to a specific lifeline facilitate the identification of functional requirements for the technical realization of individual DT and CPS elements. These functional requirements and the interaction between elements represent the main input for the technical (i.e., behavioral and structural) design of DT and CPS elements in step 7.

Since SDs can be used at any design stage, we can start to create SDs during system conception and iteratively refine these SDs during detailed system design. This refinement can be done in a structured manner by dividing elements and interactions into their sub-components and sub-functions until no further division is possible or meaningful. In [10], we provide the detailed technical design of the shown case study.

## 6 Evaluation and Discussion

Regarding the coverage of the stated research question, we can report based on our experiences and the results published in [10] that our proposed method supports designers in the derivation of requirements for CPS elements and their associated IPSS services in general and for DTs that act as IPSS service enablers in particular. Our method achieves this through the structured integration of DevOps principles into the IPSS lifecycle activities that are guided by the integrated analysis of multiple stakeholder demands. Since our method builds up on standard UML Sequence Diagrams, it can be easily integrated into a typical requirements and systems engineering landscape.

For validation, we applied our method to a representative real-world use case, i.e., for deriving the necessary requirements for realizing DT-enabled DevOps for CPS IPSS. This resulted in the publication of generic design models and their implementation and evaluation in [10]. These models and the implementation demonstrate that our method can be feasibly used for technical DT and CPS design in real world use cases.

## 7 Conclusion

This article contributes to bridging the gap that currently exists between the DT as a concept promising to enable the design of adaptive, context-aware Smart IPSS and its implementation and deployment on industrial asset and fleet levels. It does this by proposing a systematic method for deriving technical design requirements for a DT-driven CPS infrastructure that underlies essential IPSS services such as remote maintenance, control, and reconfiguration. Key insights gained through this work are that there is the need for multiple DTIs designed for specific lifecycle activities and needs. The DTI design, operation, and continuous adaptation and optimization are enabled through a DevOps-like process that is—contrary to its counterpart in the pure IT domain—highly constrained by the physical assets that are maintained, monitored, and controlled through the DTIs. At this stage, the main limitation of the presented research is a limited validation coverage due to the challenges associated with the vast scope the concept addresses (both design and operation domains) and the need for the intervention in mission-critical industrial systems for validation purposes. Our future research activities will focus on the impact that Digital Twins will have on design processes once IPSSs are designed and operated on their basis. There will also be new opportunities for specialized design methods and organizations that we plan to investigate.

## Acknowledgments

The authors thank Andritz Hydro GmbH and the Austrian Research Funding Agency FFG for supporting this research.